\documentclass[sigconf]{acmart}

\usepackage{multirow}
\usepackage{array}
\usepackage{color-edits}
\usepackage{microtype}
\usepackage{hyperref}
\usepackage{url}
\usepackage{booktabs}
\usepackage{graphicx}
\usepackage{color-edits}
\usepackage{booktabs, multirow}
\usepackage{pdflscape}
\usepackage{amsmath, amsfonts}
\usepackage{bbm}
\usepackage{nicefrac}       
\usepackage{microtype}      
\usepackage{framed}

\hypersetup{colorlinks=true, citecolor=Navy, linkcolor=blue, urlcolor=blue}

\AtBeginDocument{%
  }

\copyrightyear{2025}
\acmYear{2025}
\setcopyright{rightsretained}
\acmConference[CHI EA '25]{Extended Abstracts of the CHI Conference on Human Factors in Computing Systems}{April 26-May 1, 2025}{Yokohama, Japan}
\acmBooktitle{Extended Abstracts of the CHI Conference on Human Factors in Computing Systems (CHI EA '25), April 26-May 1, 2025, Yokohama, Japan}\acmDOI{10.1145/3706599.3706727}
\acmISBN{979-8-4007-1395-8/25/04}

\begin{document}


\title[]{Emerging Practices in Participatory AI Design in Public Sector Innovation}






\author{Devansh Saxena}
\affiliation{%
  \institution{The Information School, University of Wisconsin-Madison}
  \country{USA}
}
\email{devansh.saxena@wisc.edu}

\author{Zoe Kahn}
\affiliation{%
  \institution{School of Information, University of California, Berkeley}
  \country{USA}
}

\author{Erina Seh-Young Moon}
\affiliation{%
  \institution{Faculty of Information, University of Toronto}
  \country{Canada}
}

\author{Lauren M. Chambers}
\affiliation{%
  \institution{School of Information, University of California, Berkeley}
  \country{USA}
}
\author{Corey Jackson}
\affiliation{%
  \institution{The Information School, University of Wisconsin-Madison}
  \country{USA}
}
\author{Min Kyung Lee }
\affiliation{%
  \institution{School of Information, The University of Texas at Austin}
  \country{USA}
}
\author{Motahhare Eslami}
\affiliation{%
  \institution{Human-Computer Interaction Institute, Carnegie Mellon University}
  \country{USA}
}
\author{Shion Guha}
\affiliation{%
  \institution{Faculty of Information, University of Toronto}
  \country{USA}
}
\author{Sheena Erete}
\affiliation{%
  \institution{College of Information, University of Maryland}
  \country{USA}
}
\author{Lilly Irani}
\affiliation{%
  \institution{Department of Communication, University of California San Diego}
  \country{USA}
}
\author{Deirdre Mulligan}
\affiliation{%
  \institution{School of Information, University of California, Berkeley}
  \country{USA}
}
\author{John Zimmerman}
\affiliation{%
  \institution{Human-Computer Interaction Institute, Carnegie Mellon University}
  \country{USA}
}

\renewcommand{\shortauthors}{Devansh Saxena et al.}

\begin{abstract}
Local and federal agencies are rapidly adopting AI systems to augment or automate critical decisions, efficiently use resources, and improve public service delivery. AI systems are being used to support tasks associated with urban planning, security, surveillance, energy and critical infrastructure, and support decisions that directly affect citizens and their ability to access essential services. Local governments act as the governance tier closest to citizens and must play a critical role in upholding democratic values and building community trust especially as it relates to smart city initiatives that seek to transform public services through the adoption of AI. Community-centered and participatory approaches have been central for ensuring the appropriate adoption of technology; however, AI innovation introduces new challenges in this context because participatory AI design methods require more robust formulation and face higher standards for implementation in the public sector compared to the private sector. This requires us to reassess traditional methods used in this space as well as develop new resources and methods. This workshop will explore emerging practices in participatory algorithm design – or the use of public participation and community engagement - in the scoping, design, adoption, and implementation of public sector algorithms.
\end{abstract}




\maketitle


\newpage
\section{Introduction}

The rapid development of Artificial Intelligence (AI) and data-driven technologies create complex new contexts for
government administration and public services \cite{verhulsdonck2023smart, whitney2021hci}. On one hand, local governments want smart cities, improved public service delivery, and better decision-making \cite{verhulsdonck2023smart}. For example, AI optimizes traffic flow for congested streets, predicts maintenance needs, improves waste management and recycling, and enhances transit services \cite{zhao2021smart}. On the other hand, algorithmic tools have been used to 
`predict' where crime might happen \cite{ziosi2024evidence, haque2024we}, automate the allocation of government benefits \cite{citron2008technological}, detect the location of suspected gunfire \cite{sinha2023dangers}, automate ticketing for traffic violations \cite{sloan2020beyond}, and `modernize' working-class urban centers \cite{middha2022e-gentrification} - initiatives that perpetuate systemic inequalities and processes of minoritization along lines of race, class, and more \cite{benjamin2019race, eubanks2018automating, wang2018carceral}. These public sector systems are developed or acquired using public resources to serve the public’s interest \cite{dantec2013infrastructuring}. Therefore, it is crucial to employ participatory design (PD) methods to ensure that AI systems align with the unique needs and values of the communities they serve. Involving citizens, local stakeholders, and diverse voices in the design process will promote transparency, build trust, and help identify potential biases, harms, or unintended consequences before deployment \cite{disalvo2012communities, whitney2021hci}. This collaborative approach helps maximize AI’s benefits while minimizing its unintended harms.

Despite this growing consensus \cite{saxena2020collective, stapleton2022has, wan2023community}, there is limited evidence to suggest that PD leads to responsible AI innovation and whether PD methods reenvisioned in the AI landscape still align with their traditional roots \cite{delgado2023participatory, birhane2022power}. AI innovation introduces new challenges that traditional PD methods may not fully address, such as algorithmic transparency, data privacy concerns, and bias mitigation \cite{delgado2023participatory}. This is further complicated by the fact that PD in practice is not systematic. Practitioners create localized activities and ways of engaging stakeholders where approaches can differ in the depth of community involvement, the methods used, and the degree of influence stakeholders have on final decisions \cite{corbett2023power, delgado2023participatory}. However, this may also lead to the reappropriation of PD methods that result in “pseudo-participation” \cite{palacin2020design, birhane2022power}. Consequently, given that the strength of PD methods lies in their localized and situated approach - coupled with the complexities introduced by AI \cite{hu2024enrolling} - it is necessary to deliberate on best practices in the use of PD for responsible public sector AI innovation.

In July 2024, the U.S. Open Government Secretariat, U.S. Tech Policy Network, and the White House Office of Science and Technology Policy (OSTP) co-hosted an event on participatory algorithm design in government that featured local and federal officials and academic experts \cite{OSTP_event}. This event showcased experiences from various public sector projects where participatory practices have been implemented in algorithm design. Several organizers proposing this workshop also served as panelists, shared insights from their work, and recognized the need for deeper exploration. While progress has been made, more comprehensive discussions are necessary to address the ethical, social, and technical challenges that arise in public AI innovation projects, especially concerning the fair and ethical conceptualizing of community engagement and participation and developing a shared understanding of best practices that facilitate, shape, and capture it through and during participatory design processes.

To start addressing some of these challenges, we propose a hybrid, day-long workshop that will expand upon the OSTP event \cite{OSTP_event}, and bring together academics and practitioners to further examine and deliberate on participatory design practices. This workshop will provide a dedicated space to assess successful case studies, refine methods, and exchange best practices to ensure the design of algorithms is equitable, inclusive, and responsive to public needs. As research in participatory AI design expands, it becomes increasingly important to disentangle the work carried out in the public sector, given its distinctive constraints and challenges, from work in the private sector. Additionally, there is a need to synthesize empirical methods and to conceptualize and formalize key concepts. Without a theoretical foundation to anchor our understanding, findings from studies on enhancing public participation and community engagement may appear ad-hoc and fragmented. Lacking this synthesized understanding, the research may continue to grow in volume, but without necessarily deepening insight or delivering significant real-world impact.

\section{Workshop Themes}

Our workshop will concentrate on three key objectives in parallel to a complementary set of case studies that advanced participatory AI design in public innovation. First, we will evaluate specific methods currently employed in public sector initiatives, identifying particular strengths and weaknesses that impact community engagement. Next, we will discuss existing frameworks for participatory design of civic technologies and AI where we will conceptualize essential elements relevant to public engagement: those that enhance community involvement and identify measurable or documentable outcomes. Finally, we will explore challenges associated with public agencies’ procurement of AI services and the need for participatory requirements in vendor contracts. Below, we discuss the workshop themes in more detail.

\subsection{Evaluating Participatory Methods: Strengths, Weaknesses, and Impact on Community Engagement}
We will evaluate participatory methodologies currently used in public sector AI initiatives to assess how effectively these methods facilitate community engagement. This involves examining approaches such as citizen panels, co-design workshops, and public consultations. We will identify the strengths that enable successful collaboration, such as transparent communication and inclusive representation, and weaknesses, like lack of follow-up or insufficient diversity in stakeholder input, that may limit the impact and trustworthiness of these processes.

For example, Barcelona's \textit{Decidim} platform \cite{garcia2023collective} has been successful in engaging citizens in urban planning through a participatory, open-source digital platform that allows residents to influence policy decisions directly. A key strength of this method is its transparency and accessibility, enabling broad participation. However, it also faces challenges, such as ensuring sustained engagement and responding to conflicting interests from diverse community groups. Similarly, the City of San Antonio has organized \textit{Smart City Sandboxes} \cite{SanAntonio_event} as a way to allow citizens to interact with AI, collect public feedback, and provide access to tech-based community services. The workshop will explore these cases and others, in both local and federal government contexts.

\subsection{Defining Key Elements and Outcomes for Meaningful Public Involvement in Public Sector AI}
Next, we will focus on conceptualizing essential elements that can foster deeper community involvement in public AI initiatives as well as determining clear, measurable outcomes to assess the success of public engagement efforts. We will explore key factors that make participatory processes meaningful and impactful, such as building trust, ensuring transparency, fostering inclusivity, and creating opportunities for genuine influence on decision-making.

For instance, recent studies at CHI have found that HCI methods such as \textit{comicboarding} and the use of design probes can help scaffold early-stage conversations with community members and allow them to provide critical and detailed feedback on AI projects \cite{kuo2023understanding, haque2024we}. Participants in these studies questioned whether the technical formulation of a problem is even socially relevant, drew attention to systemic and structural issues invisible to AI developers, and helped uncover the downstream impact of such systems on their communities. These approaches helped build trust and elicit meaningful feedback, lowered the information barrier, and facilitated knowledge sharing.

To gauge the effectiveness of such participatory efforts, we will also identify measurable or documentable outcomes. Examples of simple outcomes traditionally used include the number of community members involved in the sessions, the diversity of participants, the level of public trust in the proposed technology (measured through surveys), etc. Recently, researchers have also focused on the degree to which community input influences the final design or implementation of a project. For example, \textit{Boston Smart City Playbook} \cite{gordon2023toward}, was used to measure citizen feedback incorporation evaluated through the documentation of changes made to project plans based on public input. The success was not only in participation numbers but also in documenting how feedback was directly incorporated into policy changes. By identifying these elements and setting measurable goals, we can develop methods to strengthen community engagement.


\subsection{Tackling Procurement: Embedding Participation Requirements in AI Vendor Contracts}
Finally, we will explore how public agencies can integrate participation requirements into contracts, especially considering that most agencies do not have the capacity to build AI systems in-house and often procure AI solutions from private vendors \cite{sloane2021ai}. It is essential that these vendors follow principles of community engagement, transparency, and accountability. We will discuss how specific participation guidelines in procurement contracts can be embedded that clearly outline expectations for vendors to involve communities in the design, testing, and evaluation of AI systems. This can include contractual clauses mandating public consultations or co-design sessions, such as requiring vendors to hold community workshops to gather citizen input and ensure diverse stakeholder representation in decision-making.

For instance, Amsterdam’s \textit{AI Procurement Strategy} \cite{mcbride2024towards} suggests some pathways for successful ethical guidelines that require vendors to follow the city’s transparency, data ownership, governance, and risk management requirements. Moreover, vendors must show how they will engage citizens throughout the project and ensure that AI systems address public concerns. This early requirement may help foster accountability and public trust, aligning vendor practices with community values. The strategy is still in the testing phase, as the city aims to understand how to evaluate these guidelines in practice. As another example, civil liberties groups including the ACLU have developed the Community Control Over Police Surveillance (CCOPS) campaign, which provides resources for local organizers to draft laws establishing public oversight over police surveillance technology procurement. As of 2021, at least 20 jurisdictions have enacted CCOPS ordinances, slowing law enforcement's technology acquisition and involving community leaders in evaluation processes typically hidden from the public eye \cite{southerland2023masters, irani2021oversight, benjamin2024resisting}.

\section{Pre-workshop Plans}

\subsection{Target Audience and Recruitment}
The target audience for this workshop includes both researchers and practitioners who are studying or are actively involved in the development, implementation, and evaluation of AI systems, particularly in the public sector. The participant recruitment strategy will focus on inviting practitioners from academic institutions, government agencies, NGOs, community organizations, and private sector organizations who are working in areas of public interest technology, civic AI, participatory design, and AI ethics. Outreach will be conducted via academic networks, industry associations, conferences, and targeted invitations to AI public sector experts. An open call will also be shared through social media, research forums, and practitioner groups to ensure diverse perspectives. Because this workshop is hybrid, it will be more accessible to participants from some communities.


\subsection{Paper Submissions and Review Process}
Participants are encouraged to submit a 2-3 page paper, detailing either research related to one of the workshop’s core themes or providing a motivation statement that explains their reasons for attending the workshop. We will be flexible about length and format to accommodate the norms of practitioners not from academia. These themes might include topics such as participatory methodologies, public engagement, or ethical AI in government settings. We will use EasyChair as the submission platform to ensure a streamlined and accessible process for participants.

Each submission will be reviewed by two workshop organizers, who will assess not only the quality and relevance of the work but also how it contributes to fostering an interdisciplinary and well-balanced group of participants. By focusing on diversity in expertise, background, and approach, we aim to ensure that the workshop includes voices from various fields---ranging from academia to government to industry---thereby enriching the conversation and promoting cross-disciplinary collaboration. During the reviewing process, we will especially look for new participatory approaches and practical experiences to ensure that participants reflect a broad spectrum of knowledge essential for addressing the challenges of participatory AI design in public-sector projects.

\section{Workshop Structure}

\begin{table}[]
\begin{tabular}{l|l}
\hline
\textbf{Time} & \textbf{Activity}\\
\hline
9:00 - 9:15 & Welcome \\
9:15 - 10:15 & Keynote Talk \\
10:15 - 10:30 & Coffee Break \\
10:30 - 11:30 & Paper Presentations \\
11:30 - 11:45 & Information about afternoon sessions \\
11:45 - 1:00 & Lunch\\
1:00 - 2:30 & Group Activity 1: Breakout Groups\\
2:30 - 3:00 & Coffee Break\\
3:00 - 4:30 & Group Activity 2: Design Sprint\\
4:30 - 5:00 & Closing Remarks\\
\hline
\end{tabular}
\caption{Tentative Workshop Schedule}
\label{tab:casenote_example}
\end{table}

We anticipate having between 40 and 60 attendees, based on our experience with similar workshops in the past \cite{stapleton2022has, wan2023community, saxena2020collective}. We plan to hold a one-day workshop, running from 9:00 AM to 5:00 PM local time (with breaks included), in a hybrid format. We will combine Zoom for live presentations, Slack for virtual Q\&A and discussions, and Miro board as a live, interactive document for knowledge sharing and activities throughout the workshop. From past workshops, we noticed participants appreciated the Slack and Miro board setup, where we created session-specific channels for live interactions and Miro panels for sharing resources and building a collective knowledge base. The organizing team will also collaborate with the CHI 2025 technical team to use the available streaming options, ensuring accessibility through captioning and reducing the need to switch between platforms.


The tentative workshop schedule is outlined in Table 1. To synthesize knowledge from diverse communities and encourage impactful research, ample time will be allocated for group discussions and activities in the afternoon session. This format aims to connect participants with similar interests, offering them the opportunity to contribute and learn. The morning session will open with a keynote address, followed by participants sharing their accepted work through paper presentations. These will take the form of 5-minute lightning talks, followed by a combined 10-minute Q\&A session. To minimize disruptions, these talks will be pre-recorded, ensuring that online attendees can view high-quality presentations rather than watching live streams of in-person presenters. All recordings will be made available online for asynchronous access.

In the afternoon, we plan to allocate time for two sessions of group activities. The first activity will involve participants breaking into three groups based on workshop themes (i.e., methods, measures, and participation requirements). Participants will share case studies and map successful strategies along with related constraints. In the second activity, organizers will present a design brief featuring a hypothetical AI initiative in the public sector. The breakout groups will then focus on the participatory methods needed for effective community engagement, identify measures to capture this engagement and determine how to integrate these participation requirements into binding contracts.

Each group will have at least one facilitator. We will also encourage authors of accepted papers to join groups aligned with their topics, fostering focused discussions within the broader theme, but specific to their research. For hybrid participation, activities will use collaborative platforms like Miro. Lightning talks and presentations will be hosted on Google Slides and made available on the workshop website and Miro as part of the workshop's living document.

\section{Post-workshop Plan}
After the workshop, we will engage in activities to promote knowledge sharing and community building. we will share the Miro board with the participants and the academic community. The board acts as a living knowledge repository with notes and resources shared and developed throughout the workshop activities. We will also publish all contributions on the workshop website to ensure broader dissemination and include selected papers, with authors' permission, in the workshop proceedings on \textit{arXiv}. Finally, we will share findings from the workshop, especially the group activities as Medium posts and possibly as an article in a special journal issue.

\section{Call for Participation}
We invite researchers, practitioners, and professionals working at the intersection of AI, public policy, and civic engagement to join our upcoming hybrid workshop on \textit{Emerging Practices in Participatory AI Design in Public Sector Innovation}. This workshop will explore critical themes and methods that facilitate public participation and community engagement---in the scoping, design, and implementation of public sector algorithms that are more inclusive and responsive to public needs. Specifically, we will explore the following themes at this workshop:

\textbf{Evaluating Emerging Participatory Methods.} We will assess various participatory design methods currently in use within public sector AI projects, highlighting their strengths and weaknesses in engaging communities. Participants will share insights on how these methods can be adapted or expanded to ensure more meaningful public involvement.

\textbf{Assessing Key Elements that Enhance Participation and Measurable Outcomes.} This theme focuses on identifying the essential elements that foster effective public participation in the public sector, as well as exploring measurable outcomes that can demonstrate the success of engagement efforts. We will discuss strategies for building trust, ensuring transparency, and capturing diverse perspectives necessary for equitable AI systems.

\textbf{Embedding Participation Requirements in AI Procurement Contracts.} As many cities acquire AI from private vendors, this theme will explore how public agencies can include participation requirements in procurement contracts. We will discuss how embedding these requirements can ensure accountability and alignment with public interests.

We encourage submissions in the form of 2-3 page papers, which can address any of the workshop’s themes or provide a motivation statement explaining your interest in participatory AI design. Submissions will be peer-reviewed via EasyChair by two organizers, with an emphasis on both the quality of the work and ensuring a diverse, interdisciplinary group of participants. This is an opportunity to contribute to and help shape the evolving landscape of AI in public innovation. For more details and submission guidelines, please visit our workshop website \footnote{\textbf{Workshop website} --- https://participatoryaidesign.github.io/home} or reach out to the organizers. We look forward to your participation!

\section{Organizer's Bios}

\textbf{Devansh Saxena} is an Assistant Professor in the Information School and the School of Computer, Data, and Information Sciences at the University of Wisconsin-Madison. His research focuses on participatory design and responsible AI innovation in the public sector and designing systems that center well-being, elevate human expertise, and collective decision-making. Before joining UW-Madison, he was a Presidential Postdoctoral Fellow at Carnegie Mellon University in the Human-Computer Interaction Institute where his research focused on developing new methods and tools that support AI innovation at the earliest stages of ideation, problem formulation, and project selection.

\vspace{0.1cm}
\textbf{Zoe Kahn} is a PhD Candidate at the UC Berkeley School of Information. To inform the design of more responsible tech and tech policy, her research integrates the perspectives of people with lived expertise alongside people with domain expertise (e.g., data scientists, policymakers, privacy scholars). She uses qualitative methods to understand the perspectives and experiences of impacted communities, and leverages storytelling to influence the design of technical systems and the policies that surround their use. In the process, she has developed new methods for participatory AI. Zoe has conducted empirical studies in rural villages in Togo, rural ranching communities in the United States, among tech workers at large engineering organizations, and users on Twitter.

\vspace{0.1cm}
\textbf{Erina Moon} is a PhD student at the Faculty of Information at the University of Toronto. In her research, she studies the use of and design of public sector data and AI systems. Through collaborative engagements with local governments and impacted stakeholders within homelessness and housing services, she uses mixed methods to examine how data and computational systems can augment or hinder decision-making, resource allocation, and measurement. 

\vspace{0.1cm}
\textbf{Lauren Chambers} is a PhD student at the UC Berkeley School of Information, where she studies the intersection of data, technology, and sociopolitical advocacy. Previously Lauren was the staff technologist at the ACLU of Massachusetts, where she explored government data in order to inform citizens and lawmakers about the effects of legislation and government on our civil liberties. Her current research explores the roles of 'public interest technologists' within civil society as they are shaping policy, reconfiguring service delivery, and transforming political campaigns. 

\vspace{0.1cm}
\textbf{Corey Jackson} is an Assistant Professor in the Information School at the University of Wisconsin, Madison. His research bridges human-centered computing, computer-supported cooperative work, and design epistemologies from fields such as organization studies, psychology, and education. His work employs qualitative and quantitative methodologies and is informed by a socio-technical perspective and theories. His research contributes to the design of open collaboration platforms and is directed by two complementary approaches: (1) documenting human experiences online and (2) implementing and evaluating system affordances.

\vspace{0.1cm}
\textbf{Min Kyung Lee} is an Assistant Professor in the School of Information at the University of Texas at Austin. Dr. Lee has conducted some of the first studies that empirically examine the social implications of algorithms; emerging roles in management and governance in society, looking at the impacts of algorithmic management on workers as well as public perceptions of algorithmic fairness. She has proposed a participatory framework that empowers community members to design matching algorithms for their own communities.

\vspace{0.1cm}
\textbf{Motahhare Eslami} is an Assistant Professor at the School of Computer Science, Human-Computer Interaction Institute (HCII), and Institute for Software Research (ISR), at Carnegie Mellon University. She earned her Ph.D. in Computer Science at the University of Illinois at Urbana-Champaign. Motahhare’s research goal is to investigate the existing accountability challenges in algorithmic systems and to empower the users of algorithmic systems, particularly those who belong to marginalized communities or those whose decisions impact marginalized communities, make transparent, fair, and informed decisions in interaction with algorithmic systems.

\vspace{0.1cm}
\textbf{Shion Guha} is an Assistant Professor in the Faculty of Information and cross-appointed to the Department of Computer Science at the University of Toronto. His research interests include human-computer interaction, data science, and public policy. He’s been involved in developing the field of Human-Centred Data Science. This intersectional research area combines technical methodologies with interpretive inquiry to address biases and structural inequalities in socio-technical systems. He is the author of Human-Centered Data Science: An Introduction, an Amazon Best Selling textbook published by MIT Press in 2022.  Shion wants to understand how algorithmic decision-making processes are designed, implemented, and evaluated in public services. In doing so, he often works with marginalized and vulnerable populations, such as communities impacted by child welfare, homelessness, and public health systems. His work has been supported by grants from the Canadian Institute for Advanced Research, the National Science and Engineering Research Council, the National Science Foundation, the American Political Science Association, Meta, IBM, etc. He has been featured in the media (Newsweek, Associated Press, ACLU, ABC, NBC, Gizmodo, etc.) 

\vspace{0.1cm}
\textbf{Sheena Erete} is an Associate Professor in the College of Information at the University of Maryland, College Park, the associate director of research for the Artificial Intelligence Interdisciplinary Institute at Maryland (AIM), and the director of the Community Research, Equity, and Design Collective (CREED) Collective. Her research focuses on co-designing socio-cultural technologies, practices, and policies with historically marginalized populations to amplify their existing efforts in addressing social issues. Centering community knowledge and strengths, she leverages Black feminist epistemologies and transformative justice practices to address issues such as equity in AI/ML tools, community safety, education, political efficacy, and economic development.

\vspace{0.1cm}
\textbf{Lilly Irani} is an Associate Professor of Communication \& Science Studies at the University of California, San Diego. She also serves as faculty in the Design Lab, Institute for Practical Ethics, the program in Critical Gender Studies, and sits on the Academic Advisory Board of AI Now (NYU). She is the author of Chasing Innovation: Making Entrepreneurial Citizens in Modern India (Princeton University Press, 2019) and Redacted (with Jesse Marx) (Taller California, 2021). Chasing Innovation has been awarded the 2020 International Communication Association Outstanding Book Award and the 2019 Diana Forsythe Prize for feminist anthropological research on work, science, or technology, including biomedicine. She is also a co-founder of the digital worker advocacy organization Turkopticon. Her work has appeared at ACM SIGCHI, New Media \& Society, Science, Technology \& Human Values, South Atlantic Quarterly, and other venues. She sits on the Steering Committee of the San Diego Transparent and Responsible Use of Surveillance Technology Coalition and also serves on the board of United Taxi Workers San Diego.

\vspace{0.1cm}
\textbf{Deirdre K. Mulligan} is a Professor in the School of Information at UC Berkeley, a faculty Director of the Berkeley Center for Law \& Technology, a co-organizer of the Algorithmic Fairness \& Opacity Working Group, an affiliated faculty on the Hewlett funded Berkeley Center for Long-Term Cybersecurity and a faculty advisor for it's AI Policy Hub, and a faculty advisor to the CITRIS Policy Lab. Mulligan’s research explores legal and technical means of protecting values such as privacy, freedom of expression, and fairness in socio-technical systems.  Mulligan served as Principal Deputy U.S. Chief Technology Officer at the White House Office of Science and Technology Policy, and Director of the National Artificial Intelligence Initiative Office (NAIIO), in the Biden-Harris Administration. At OSTP, Mulligan led the Technology Team that works to advance technology and data to benefit all Americans. Under her leadership, the Tech Team leveraged technology and data to equitably deliver services, brought technology and data expertise to federal policy formation and implementation, and ensured that America led the world in values-driven technological research and innovation.

\vspace{0.1cm}
\textbf{John Zimmerman} is the Tang Family Professor of AI and HCI at HCI Institute within Carnegie Mellon’s School of Computer Science. He researches and designs human-AI interaction, human-robot interaction, and methods of innovating AI products and services. For more than twenty years, Professor Zimmerman has designed novel, intelligent systems ranging from one of the first TV show recommenders to a crowd-sourced, transit arrival system to a decision support tool for implanting mechanical hearts to a system that keeps parents from forgetting to pick up their children. He has published more than 150 papers and is a member of the ACM CHI Academy. He teaches courses in service design, lean startup, and the design of AI products and services. While working for Philips, he invented the way everyone scrolls on their smartphone.


\newpage
\clearpage
\newpage
\bibliographystyle{ACM-Reference-Format}
\bibliography{refs}

\end{document}